\documentclass[aps,prb,showpacs,groupedaddress,amssymb,twocolumn]{revtex4}

\usepackage{bm,amsmath}
\usepackage[dvips]{graphicx,color}

\begin{document}

\title{Thermodynamic equivalence of certain ideal Bose and Fermi 
gases}

\author{Kelly R. Patton and Michael R. Geller}
\affiliation{Department of Physics and Astronomy, University of 
Georgia, Athens, Georgia 30602-2451}
\author{Miles P. Blencowe}
\affiliation{Department of Physics and Astronomy, Dartmouth College, 
Hanover, New Hampshire 03755 }

\date{June 11, 2004}

\begin{abstract}
We show that the recently discovered thermodynamic ``equivalence'' 
between noninteracting Bose and Fermi gases in two dimensions, and 
between one-dimensional Bose and Fermi systems with linear 
dispersion, both in the grand-canonical ensemble, are special cases 
of a larger class of equivalences of noninteracting systems having an 
energy-independent single-particle density of states. We also 
conjecture that the same equivalence will hold in the grand-canonical 
ensemble for any noninteracting quantum gas with a discrete 
ladder-type spectrum whenever $\sigma \Delta  / N k_{\rm B} T$ is 
small, where $N$ is the average particle number and $\sigma$ its 
standard deviation, $\Delta$ is the level spacing, $k_{\rm B}$ is 
Boltzmann's constant, and $T$ is the temperature. 
\end{abstract}


\maketitle
\clearpage

\section{INTRODUCTION}

There has been considerable recent interest in surprising 
thermodynamic ``equivalences'' between certain ideal Bose and 
spinless Fermi gas systems, including nonrelativistic free particles 
in two dimensions,\cite{Lee} and between one-dimensional particles 
with a linear, sound-like dispersion relation.\cite{Pathria} Both 
results are valid in the grand-canonical ensemble, and assert that 
the Helmholtz free energies $F_{\rm F}$ and $F_{\rm B}$ of the Fermi 
and Bose systems, respectively, are simply related by
\begin{equation}
F_{\rm F}(T,V,N) - F_{\rm B}(T,V,N) =  {N^2 \over 2 {\cal C}},
\label{equivalence}
\end{equation}
where ${\cal C}$ is a constant that may depend on the system volume 
$V$, but is independent of temperature $T$ and the mean particle 
number $N$. The Bose and Fermi systems are assumed to have identical 
single-particle Hamiltonians, and have the same $T$, $V$, and $N$.

There are several immediate consequences of Eq.~(\ref{equivalence}), 
including:

\begin{enumerate}

\item[a.] The entropies  of the Fermi and Bose systems are identical,
\begin{equation}
S_{\rm F}(T,V,N) = S_{\rm B}(T,V,N);
\end{equation}

\item[b.] Their internal energies differ by a temperature-independent 
constant, namely
\begin{equation}
U_{\rm F}(T,V,N) - U_{\rm B}(T,V,N) =  {N^2 \over 2 {\cal C}};
\end{equation}

\item[c.] The constant-volume heat capacities are identical;

\item[d.] The chemical potentials are simply shifted by a 
temperature-independent constant,
\begin{equation}
\mu_{\rm F}(T,V,N) - \mu_{\rm B}(T,V,N) =  {N\over {\cal C}};
\end{equation}

\item[e.] The thermodynamic potentials are connected by a relation 
opposite to (\ref{equivalence}),
\begin{equation}
\Omega_{\rm F}(T,V,N) - \Omega_{\rm B}(T,V,N) =  -{N^2 \over 2{\cal 
C}};
\end{equation}

\item[f.] The pressures of the Fermi and Bose gases satisfy
\begin{equation}
P_{\rm F}(T,V,N) - P_{\rm B}(T,V,N) =  {N^2 \over 2 {\cal C} V},
\end{equation}
and again differ only by a temperature-independent constant.
\end{enumerate}
These results explain and considerably extend isolated thermodynamic 
relations that were discovered some time ago by May.\cite{May}

In this paper we demonstrate that the equivalence defined in 
Eq.~(\ref{equivalence}) holds whenever the single-particle density of 
states (DOS) is independent of energy, for which the systems 
considered in Refs.~\onlinecite{Lee, Pathria, May} are special cases. 
We then extend recent results of Schmidt and Schnack\cite{Schmidt and 
Schnack 1998} and of Crescimanno and Landsberg\cite{Crescimanno and 
Landsberg} on harmonically confined Bose and Fermi gases in one 
dimension to further enlarge the class of equivalences to include any 
ideal quantum gas with a discrete ladder-type spectrum, in the limit
\begin{equation}
{\sigma \Delta   \over N k_{\rm B} T} \rightarrow 0.
\label{limit condition}
\end{equation}
Here $N$ is the mean particle number, $\sigma$ is the standard 
deviation in particle number about $N$, $\Delta$ is the energy-level 
spacing, $k_{\rm B}$ is Boltzmann's constant, and $T$ is the 
temperature. The condition (\ref{limit condition}) requires that 
either (i) the particle number is conserved, (ii) the number of 
particles is very large, (iii) the temperature is very high, (iv) the 
spectrum is continuous, or any combination of these possibilities.

\section{THERMODYNAMIC EQUIVALENCE AND THE DENSITY OF STATES}
\label{constant DOS section}

We begin by writing the grand-canonical partition function of an 
arbitrary noninteracting Bose or Fermi system as
\begin{equation}
Z = \prod_\alpha \sum_{N_{\! \alpha}} e^{- \beta (\epsilon_\alpha - 
\mu) N_{\! \alpha}}.
\label{partition function definition}
\end{equation}
Here $\alpha$ labels the quantum states of a single Bose or spinless 
Fermi particle with spectrum $\epsilon_\alpha$, and $\beta \equiv 
1/k_{\rm B} T$. The occupation numbers $N_\alpha$ take the values 
$N_\alpha = 0, 1, 2, \cdots$ for bosons and $N_\alpha = 0, 1$ for 
fermions. The thermodynamic potential $\Omega \equiv F - \mu N$ is 
given by
\begin{equation}
\Omega = - {1 \over \beta} \ln Z.
\end{equation}

Because the average number of particles is required to be the same for the 
Bose and Fermi cases, their chemical potentials $\mu$ in Eq.~(\ref{partition function definition}) 
are different. The relations between 
$\mu_{\rm B}$, $\mu_{\rm F}$, and $N$, are determined by
\begin{equation}
\sum_\alpha n_{\rm B}(\epsilon_\alpha - \mu_{\rm B}) = \sum_\alpha 
n_{\rm F}(\epsilon_\alpha - \mu_{\rm F}) = N,
\label{particle number condition}
\end{equation}
where
\begin{equation}
n_{\rm B}(x) \equiv {1 \over e^{\beta x} - 1} \ \ \ \ {\rm and} \ \ \ 
\ n_{\rm F}(x) \equiv {1 \over e^{\beta x} + 1}
\end{equation}
are the Bose and Fermi distribution functions.

Next we define a single-particle DOS according to
\begin{equation}
g(\epsilon) \equiv \sum_\alpha \delta(\epsilon - \epsilon_\alpha),
\end{equation}
which gives the number of energy levels per unit energy, as a 
function of $\epsilon$. In a translationally invariant system, 
$g(\epsilon)$ scales linearly with system volume $V$. In terms of the 
DOS, we have
\begin{equation}
\Omega_{\rm B} = {1 \over \beta} \int_{-\infty}^\infty \! d\epsilon 
\, g(\epsilon) \, \ln(1 - e^{-\beta \epsilon} z_{\rm B}) 
\label{general bose partition function}
\end{equation}
and
\begin{equation}
\Omega_{\rm F} = - {1 \over \beta} \int_{-\infty}^\infty \! d\epsilon 
\, g(\epsilon) \, \ln(1 + e^{-\beta \epsilon} z_{\rm F}),
\label{general fermi partition function}
\end{equation}
where $z_{\rm B} \equiv e^{\beta \mu_{\rm B}}$ and $z_{\rm F} \equiv 
e^{\beta \mu_{\rm F}}$ are the Bose and Fermi fugacities. 
Furthermore, condition (\ref{particle number condition}) can be 
written as
\begin{equation}
\int_{-\infty}^\infty \! d\epsilon \, { g(\epsilon) \over e^{\beta 
\epsilon} \, z_{\rm B}^{-1} - 1} = 
\int_{-\infty}^\infty \! d\epsilon \, { g(\epsilon) \over e^{\beta 
\epsilon} \, z_{\rm F}^{-1} + 1}.
\label{new particle number condition}
\end{equation}
The expressions (\ref{general bose partition function}), 
(\ref{general fermi partition function}), and (\ref{new particle 
number condition}), are valid for any
noninteracting quantum gas.

We assume that spectrum is bounded from below, and that the DOS is a 
constant, ${\cal C}$, independent of energy, above that minimum. 
Without loss of generality we can take the minimum to be at $\epsilon 
\! = \! 0$. Then
\begin{equation}
g(\epsilon) = {\cal C} \Theta(\epsilon),
\label{constant DOS}
\end{equation}
where $\Theta(\epsilon)$ is the unit step function. The most common 
example of a DOS of the form (\ref{constant DOS}) occurs for free 
nonrelativistic particles of mass $m$ in two dimensions, in which case
\begin{equation}
{\cal C} = {2m \over \pi \hbar^2}.
\end{equation}
However, there are other situations where (\ref{constant DOS}) holds 
as well, including noninteracting particles moving in one dimension 
with a linear dispersion $\epsilon(k) \propto |k|$, and also for 
particles moving in three dimensions with qubic dispersion 
$\epsilon(k) \propto |k|^3$. These cases were noted earlier by 
Pathria.\cite{Pathria} Furthermore, we note that the equivalence 
(\ref{equivalence}) would {\it not} apply to two-dimensional systems 
moving in the potential of a corrugated surface or to ideal lattice 
gas models. 

Assuming (\ref{constant DOS}), we can immediately obtain
\begin{equation}
\Omega_{\rm B} = -{{\cal C} \over \beta^2} \, {\rm Li}_2 (z_{\rm B}) 
\end{equation}
and
\begin{equation}
\Omega_{\rm F} = {{\cal C} \over \beta^2} \, {\rm Li}_2 (-z_{\rm F}),
\end{equation}
where $z_{\rm B} = 1 - e^{-\beta N/{\cal C}}$ and $z_{\rm F} = 
e^{\beta N/{\cal C}} - 1$. Here ${\rm Li}_2(x)$ is Euler's 
dilogarithm function. Furthermore, from Eq.~(\ref{new particle number condition}),
we have $\mu_{\rm F}- \mu_{\rm B} =  N / {\cal C}$ and $z_{\rm F} = z_{\rm B} /(1 - z_{\rm B})$. 
These relations, along with the identity
\begin{equation}
{\rm Li}_2(x) + {\rm Li}_2({\textstyle{x \over x-1}}) = - 
{\textstyle{1 \over 2}} [\ln(1-x)]^2 ,
\end{equation}
directly lead to the equivalence stated in (\ref{equivalence}). The 
thermodynamic equivalence evidently applies to any noninteracting 
quantum gas with a constant DOS.

It is also instructive to directly demonstrate the equivalence of the entropies: 
For a system with a constant DOS of the form (\ref{constant DOS}), the Bose
and Fermi entropies are\cite{Landau}
\begin{eqnarray}
    S_{\rm B}&=&-{\cal C}k_{\rm B}\int_{0}^{\infty} \! \! d\epsilon \, 
    \bigg[ n_{\rm B}(\epsilon \! - \! \mu_{\rm B}) \ln \! \big[n_{\rm B}(\epsilon \! - \! \mu_{\rm B})\big] \nonumber \\
    &-&  \big[1+n_{\rm B}(\epsilon \! - \! \mu_{\rm B})\big] \ln \! \big[1+n_{\rm B}(\epsilon \! - \! \mu_{\rm B})\big] \bigg]
    \label{bose entropy}
\end{eqnarray}
and
\begin{eqnarray}
    S_{\rm F}&=&-{\cal C}k_{\rm B}\int_{0}^{\infty}\! \! d\epsilon \, 
    \bigg[  n_{\rm F}(\epsilon \! - \! \mu_{\rm F}) \ln \! \big[ n_{\rm F}(\epsilon \! - \! \mu_{\rm F}) \big] \nonumber \\
   &+&  \big[1-n_{\rm F}(\epsilon \! - \! \mu_{\rm F})\big] \ln \! \big[1-n_{\rm F}(\epsilon \! - \! \mu_{\rm F})\big] \bigg]. 
    \label{fermi entropy}
\end{eqnarray}
Changing the integration variable in the Bose case to $w=e^{{\beta}(\epsilon-\mu_{B})}-1$, and in the Fermi case
to $w=e^{{\beta}(\epsilon-\mu_{F})},$ leads to
\begin{equation}
    S_{\rm B}={\cal C}k_{\rm B}^{2}T\int_{z_{\rm B}^{-1}-1}^{\infty}\! \! dw \, \bigg[ {\ln(1+w) \over w} -{\ln w \over 1+w} \bigg]
    \label{bose entropy 2}
\end{equation}
and
\begin{equation}
    S_{\rm F}={\cal C}k_{\rm B}^{2}T\int_{z_{\rm F}^{-1}}^{\infty}\! \! dw \, \bigg[ {\ln(1+w) \over w} -{\ln w \over 1+w} \bigg].
    \label{fermi entropy 2}
\end{equation}
Notice that the statistics dependence enters only in the lower 
integration limits. Because the average particle numbers are the same, 
these lower limits coincide and thus the Fermi and Bose entropies are identical.

\section{1D QUANTUM GASES IN HARMONIC POTENTIALS AND AN EXTENDED 
EQUIVALENCE}

\begin{figure}
\includegraphics[width=8.0cm]{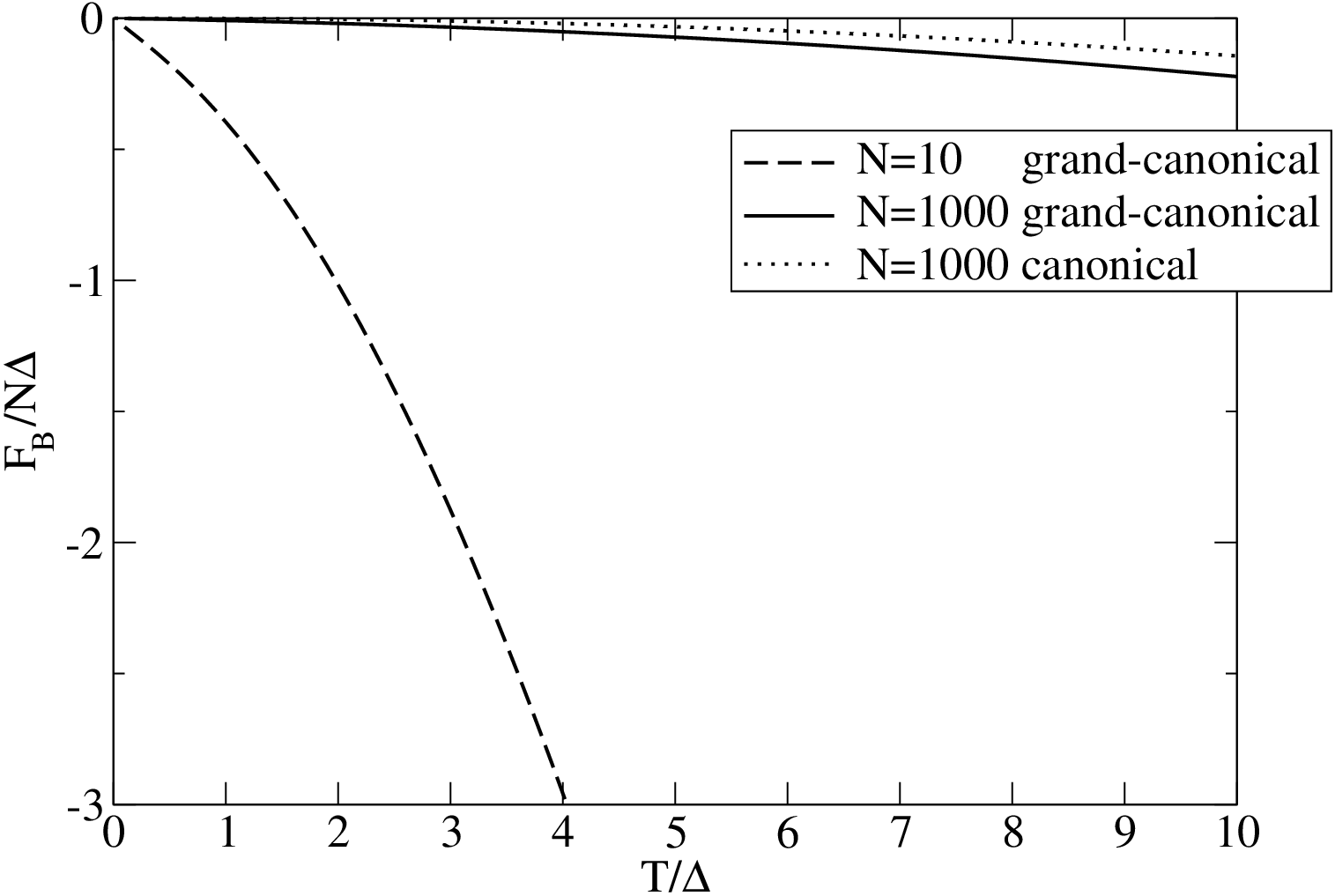}
\caption{\label{bose figure}Helmholtz free energy per particle for a 
1D ideal Bose gas in a harmonic confining potential with level 
spacing $\Delta$.}
\end{figure}

It is interesting to consider whether a constant DOS is {\it 
necessary} for the equivalence defined in Eq.~(\ref{equivalence}). In 
particular, does it apply to a system with a discrete ladder-type 
spectrum of the form
\begin{equation}
\epsilon_n =  n \, \Delta, \ \ \ \ \ n=0,1,2,3, \dots,
\label{oscillator spectrum}
\end{equation}
where $\Delta$ is the level spacing, which reduces to the case 
considered in Sec.~\ref{constant DOS section} in the limit $\Delta 
\rightarrow 0$? In this section we demonstrate that the equivalence 
does still hold, in the grand-canonical ensemble, for noninteracting 
gases with the spectrum (\ref{oscillator spectrum}), in the limit $N 
\rightarrow \infty$, and also in the canonical ensemble for any $N$. 

\begin{figure}
\includegraphics[width=8.0cm]{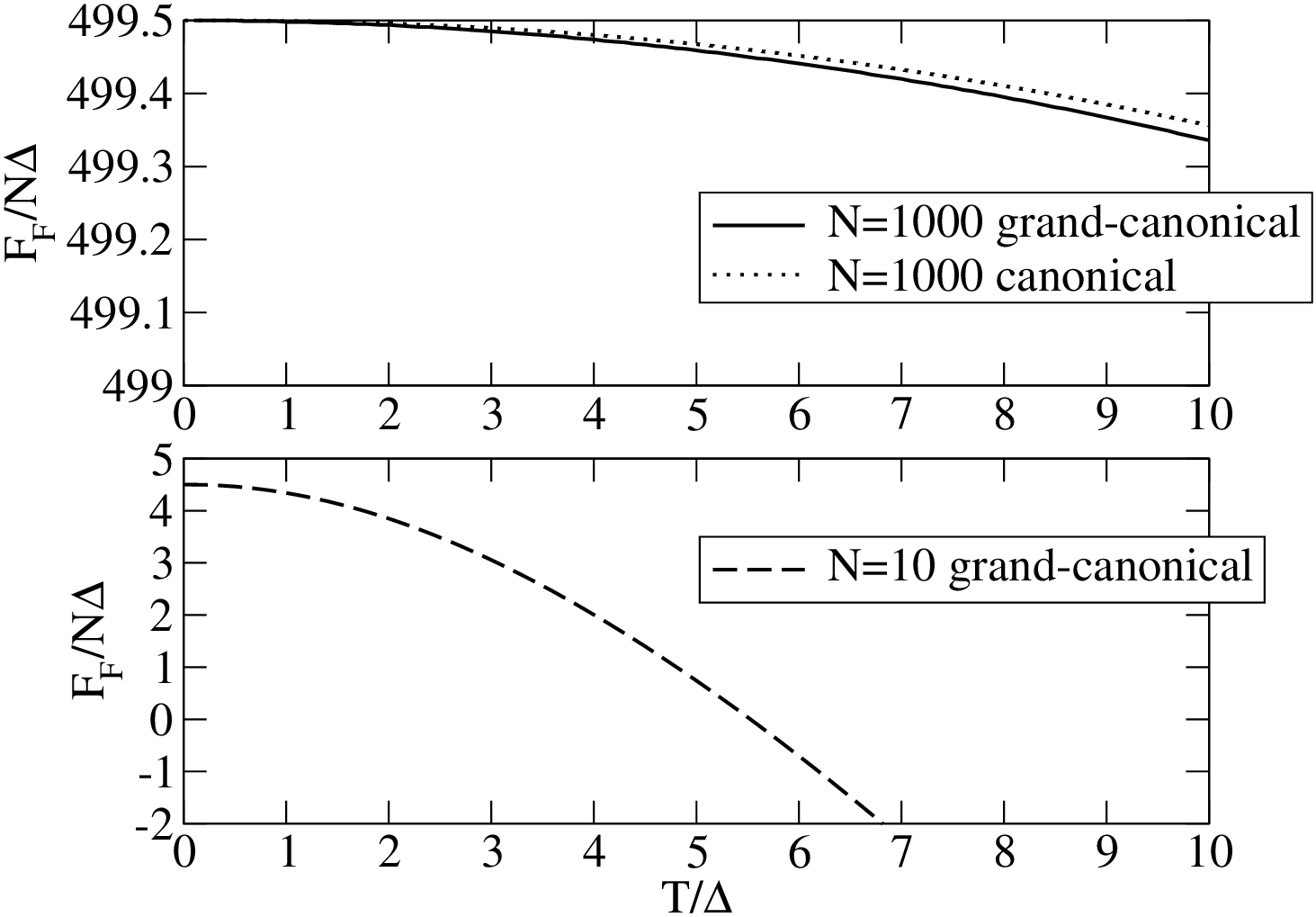}
\caption{\label{fermi figure}Helmholtz free energy per particle for a 
1D ideal Fermi gas in a harmonic confining potential with level 
spacing $\Delta$.}
\end{figure}

\begin{figure}
\includegraphics[width=8.0cm]{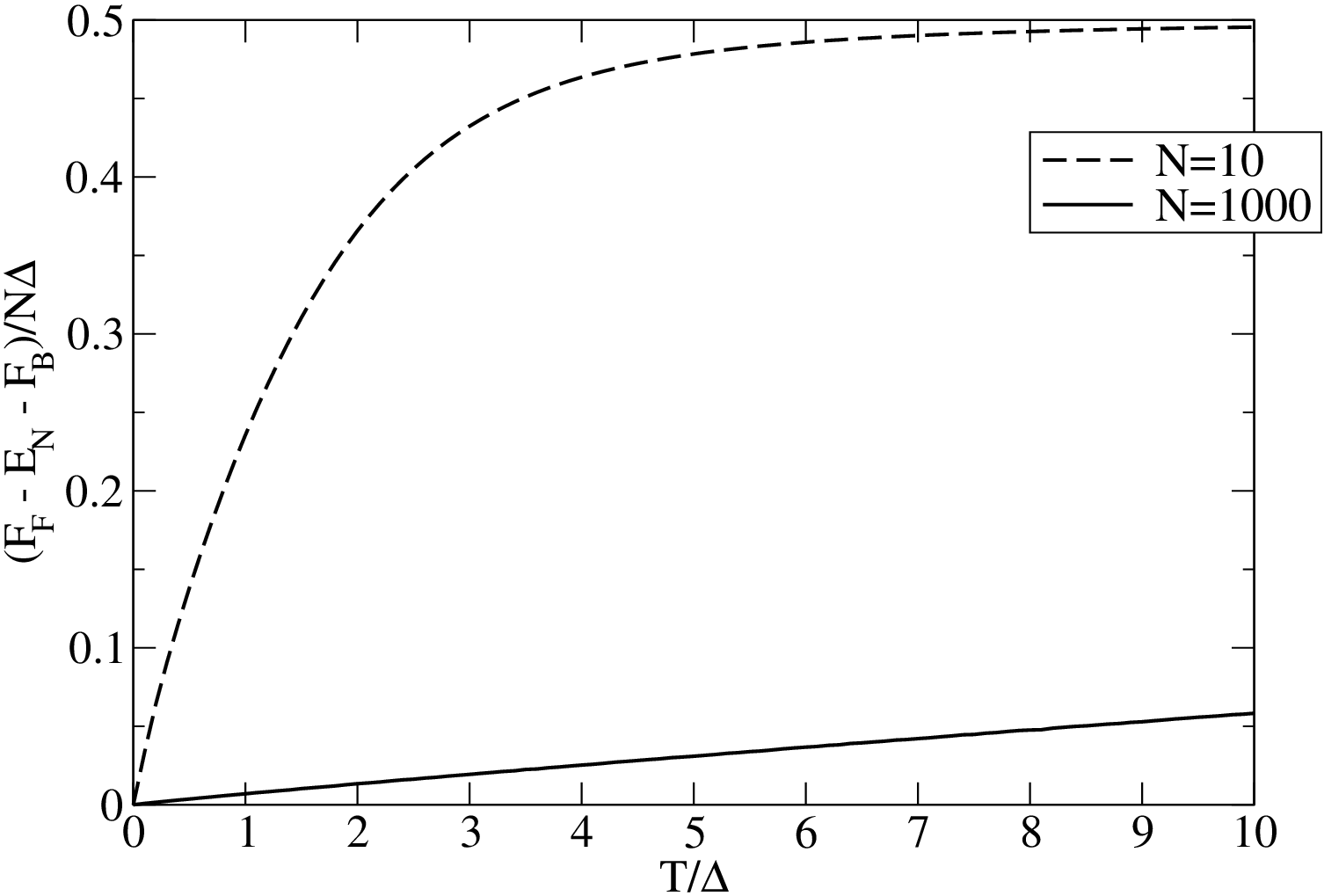}
\caption{\label{difference figure}Free energy difference per particle 
for ideal quantum gases in a harmonic confining potential. Here 
$E_{\! {\scriptscriptstyle N}}$ is the ground-state energy of the 
Fermi system. Not shown is the difference in canonical ensemble case, 
which is exactly zero for any $N$.}
\end{figure}

In Figs.~\ref{bose figure} and \ref{fermi figure} we show the 
Helmholtz free energy per particle, numerically calculated in the 
grand-canonical ensemble, for 1D quantum gases in a harmonic 
potential with level spacing $\Delta$. The free energies and 
temperatures are plotted in units of $\Delta$. In these figures, the 
solid curves are for $N=1000$ and the dashed curves are for $N=10$. 
In Fig.~\ref{difference figure}, the difference between the Fermi and 
Bose free energies are given as a function of temperature, with 
$F_{\rm F}$ shifted by the Fermi ground-state energy $E_{\! 
{\scriptscriptstyle N}}$ [defined below in Eq.~(\ref{ground state 
energy})] for convenience.

For small $N$ (dashed curve in Fig.~\ref{difference figure}), the 
free energies clearly do not differ by a temperature-independent 
constant. However, as $N$ becomes larger (solid curve), the 
equivalence does apply. These numerical results suggest that when 
particle-number fluctuations become negligible, the equivalence 
holds. To establish this result, we use the fact that in the 
large-$N$ limit, the grand-canonical free energy approaches the 
canonical free energy (shown in  Figs.~\ref{bose figure} and 
\ref{fermi figure} as dotted curves), and that the equivalence holds 
exactly in the canonical ensemble.

Writing the grand-canonical partition function $Z$ in terms of the 
canonical partition functions $Z_{\! {\scriptscriptstyle N}}$ 
according to
\begin{equation}
Z = \sum_{N=1}^\infty Z_{\! {\scriptscriptstyle N}} z^{\! 
{\scriptscriptstyle N}}, \ \ \ \ \ z \equiv e^{\beta \mu},
\end{equation}
where $z$ is the fugacity, leads to
\begin{equation}
Z_{\! {\scriptscriptstyle N}} = {1 \over N!}  \bigg( 
{\partial^{{\scriptscriptstyle N}} \! Z \over \partial z^{\! 
{\scriptscriptstyle N}}} \bigg)_{\! \! z=0}.
\end{equation}
For Bose $(\zeta \! = \! 1)$ and Fermi $(\zeta \! = \! -1)$ particles 
with spectrum (\ref{oscillator spectrum}), the grand-canonical 
partition function is
\begin{equation}
Z = \exp\bigg[\! - \zeta \sum_{n=0}^\infty \ln \big(1-\zeta b^n z 
\big)\bigg] \ \ \ \ {\rm with} \ \ \ \ b \equiv e^{-\beta \Delta},
\end{equation}
from which we obtain
\begin{equation}
Z_{\! {\scriptscriptstyle N}} = e^{-\beta E_{\! N}} \times 
\prod_{j=1}^N \bigg({1 \over 1 - b^j}\bigg).
\label{canonical Z}
\end{equation}
Here
\begin{equation}
E_{\! {\scriptscriptstyle N}}  \equiv \begin{cases} 0 & \text{for 
bosons} \\ 
{ N(N-1) \over 2} \Delta &  \text{for fermions} \end{cases}
\label{ground state energy}
\end{equation}
is the ground-state energy of $N$ particles. The result in 
Eq.~(\ref{canonical Z}) was also obtained by Schmidt and 
Schnack\cite{Schmidt and Schnack 1998} using related methods. 

According to Eq.~(\ref{canonical Z}), the canonical free energy for 
$N$ particles is simply
\begin{equation}
F_{\! {\scriptscriptstyle N}} = E_{\! {\scriptscriptstyle N}}  + 
k_{\rm B} T \sum_{j=1}^N \ln\big(1 - e^{-\beta \Delta j} \big),
\label{canonical F}
\end{equation}
the Bose and Fermi cases simply differing by the constant $E_{\! 
{\scriptscriptstyle N}}$. The second term in (\ref{canonical F}) does 
not depend on the quantum statistics parameter $\zeta$. Schmidt and 
Schnack\cite{Schmidt and Schnack 1998} also recognized the partial 
equivalence between 1D ideal Bose and Fermi gases in harmonic 
confining potentials. Later, Crescimanno and 
Landsberg\cite{Crescimanno and Landsberg} showed that the physical 
origin of that equivalence in the canonical ensemble is the exact 
mapping between the many-particle excitation spectra of both systems.

Finally, we note that the thermodynamic equivalence with the spectrum 
(\ref{oscillator spectrum}) also trivially holds in the $T 
\rightarrow \infty$ limit, because in this limit the systems become 
classical. Therefore we conjecture that the equivalence defined in 
Eq.~(\ref{equivalence}) will hold in the grand-canonical ensemble for 
any noninteracting quantum gas with a discrete ladder-type spectrum, 
whenever the quantity
\begin{equation}
{\sigma \Delta  \over  N k_{\rm B} T}
\end{equation}
is small, where $N$ is the average particle number and $\sigma$ is 
its standard deviation about $N$. This ratio roughly characterizes 
the magnitude of energy fluctuations caused by the exchange of 
particles with the environment---if allowed---relative to the thermal 
energy. We conclude that the (partial) thermodynamic equivalence 
discovered by Lee\cite{Lee} holds for ideal quantum gases with the 
spectrum (\ref{oscillator spectrum}) whenever (i) the number of 
particles is strictly conserved, (ii) the number of particles becomes 
very large so that $\sigma/N \rightarrow 0$, (iii) $T \rightarrow 
\infty$, (iv) $\Delta \rightarrow 0$, or when any combination of 
these criteria are fulfilled.

\acknowledgements

This work was supported by the  National Science Foundation under NIRT
Grant No.~CMS-0404031. MRG also acknowledges support from the National 
Science Foundation under CAREER Grant No.~DMR-0093217, and from the 
Research Corporation. We are grateful to Howard Lee for stimulating our interest 
in this problem and for useful discussions.

\end{document}